\documentclass[final,1p,times,twocolumn]{elsarticle}
\usepackage{graphicx,amssymb}
\journal{Physics Letters B}
\begin{document}
\begin{frontmatter}

\title{Evolution of the fine-structure constant in runaway dilaton models}
\author[inst1,inst2]{C. J. A. P. Martins\corref{cor1}}\ead{Carlos.Martins@astro.up.pt}
\author[inst3]{P. E. Vielzeuf}\ead{pvielzeuf@ifae.es}
\author[inst4]{M. Martinelli}\ead{martinelli@thphys.uni-heidelberg.de}
\author[inst5]{E. Calabrese}\ead{erminia.calabrese@astro.ox.ac.uk}
\author[inst6]{S. Pandolfi}\ead{stefania@dark-cosmology.dk}
\address[inst1]{Centro de Astrof\'{\i}sica, Universidade do Porto, Rua das Estrelas, 4150-762 Porto, Portugal}
\address[inst2]{Instituto de Astrof\'{\i}sica e Ci\^encias do Espa\c co, CAUP, Rua das Estrelas, 4150-762 Porto, Portugal}
\address[inst3]{Institut de F\'{\i}sica d'Altes Energies, Universitat Aut\`onoma de Barcelona, E-08193 Bellaterra (Barcelona), Spain}
\address[inst4]{Institute for Theoretical Physics, University of Heidelberg, Philosophenweg 16, 69120, Heidelberg, Germany}
\address[inst5]{Sub-department of Astrophysics, University of Oxford, Keble Road, Oxford OX1 3RH, UK}
\address[inst6]{Dark Cosmology  Centre,  Niels  Bohr  Institute,  University  of  Copenhagen,  Juliane  Maries  Vej  30,  DK-2100  Copenhagen, Denmark}
\cortext[cor1]{Corresponding author}

\begin{abstract}
We study the detailed evolution of the fine-structure constant $\alpha$ in the string-inspired runaway dilaton class of models of Damour, Piazza and Veneziano. We provide constraints on this scenario using the most recent $\alpha$ measurements and discuss ways to distinguish it from alternative models for varying $\alpha$. For model parameters which saturate bounds from current observations, the redshift drift signal can differ considerably from that of the canonical $\Lambda$CDM paradigm at high redshifts. Measurements of this signal by the forthcoming European Extremely Large Telescope (E-ELT), together with more sensitive $\alpha$ measurements, will thus dramatically constrain these scenarios.
\end{abstract}

\begin{keyword}
Cosmology \sep Dynamical dark energy \sep Fine-structure constant \sep Astrophysical observations
\end{keyword}

\end{frontmatter}

\section{Introduction}

The observational evidence for cosmic acceleration, first inferred from the luminosity distance of type Ia supernovae in 1998 \cite{SN1,SN2}, opened a new avenue in cosmological research. The most obvious task in this endeavor is to identify the source of this acceleration---the so-called Dark Energy---and in particular to ascertain whether it is due to a cosmological constant or to a new dynamical degree of freedom. While the former option, corresponding to the canonical $\Lambda$CDM paradigm, is arguably the simplest, many alternative models have been proposed and still have to be tested \cite{Dark}.

The most natural way to model dynamical energy is through a scalar field, of which the recently discovered Higgs is the obvious example \cite{ATLAS,CMS}. String theory predicts the presence of a scalar partner of the spin-2 graviton, the dilaton, hereafter denoted $\phi$. Here, we will study the cosmological consequences of a particular class of string-inspired models, the runaway dilaton scenario of Damour, Piazza and Veneziano \cite{DPV1,DPV2}. In this scenario, which among other things provides a way to reconcile a massless dilaton with experimental data, the dilaton decouples while cosmologically attracted towards infinite bare coupling, and the coupling functions have a smooth finite limit
\begin{equation}\label{eq:coupfunc}
B_i(\phi)=c_i+{\cal O}(e^{-\phi})\,.
\end{equation}
As discussed in \cite{DPV2}, provided there's a significant (order unity) coupling to the dark sector, the runaway of the dilaton towards strong coupling may yield violations of the Equivalence Principle and variations of the fine-structure constant $\alpha$ that are potentially measurable.

More than a decade after the original analysis the available $\alpha$ measurements have improved substantially \cite{LP1,LP3}, and it's therefore timely to revisit these models. Additional gains in sensitivity will be provided by forthcoming facilities such as the E-ELT: its high-resolution ultra-stable spectrograph (HIRES) will significantly improve tests of the stability of fundamental couplings and will also be sensitive enough to carry out a first measurement of the redshift drift deep in the matter-dominated era \cite{Liske,Hires}. The combination of both types of measurements is a powerful probe of dynamical dark energy, as it can distinguish between models that are indistinguishable at low redshifts \cite{Codex}. In what follows we obtain constraints on this runaway dilaton scenario using current $\alpha$ data, and also discuss how they may be further improved.

\section{Runaway dilaton cosmology}

As discussed in \cite{DPV1,DPV2}, the Einstein frame Lagrangian for this class of models is
\begin{equation}\label{eq:lagr}
{\cal L}=\frac{R}{16\pi G}-\frac{1}{8\pi G}\left(\nabla\phi\right)^2-\frac{1}{4}B_F(\phi)F^2+... \,.
\end{equation}
where $R$ is the Ricci scalar and $B_F$ is the gauge coupling function. From this one can show \cite{DPV2} that the corresponding Friedmann equation, relating the Hubble parameter, $H$, to the dilaton and the other components of the universe is as follows
\begin{equation}\label{eq:friedmann}
3H^2=8\pi G\sum_i \rho_i+H^2\phi'^2\,,
\end{equation}
where the sum is over the components of the universe, except the kinetic part of the dilaton field which is described by the last term (where the prime is the derivative with respect to the logarithm of the scale factor). The sum does include the potential part of the scalar field; the total energy density and pressure of the field are
\begin{equation}\label{phidens}
\rho_\phi=\rho_k+\rho_v=\frac{(H\phi')^2}{8\pi G}+V(\phi)
\end{equation}
\begin{equation}\label{phipres}
p_\phi=p_k+p_v=\frac{(H\phi')^2}{8\pi G}-V(\phi)\,;
\end{equation}
here $k$ and $v$ correspond to the kinetic and potential parts of the field, with the latter providing the dark energy. On the other hand, the evolution equation for the scalar field is
\begin{equation}\label{eq:field}
\frac{2}{3-\phi'^2}\phi''+\left(1-\frac{p}{\rho}\right)\phi'=-\sum_i\alpha_i(\phi)\frac{\rho_i-3p_i}{\rho}\,.
\end{equation}
Here $p=\sum_ip_i$, $\rho=\sum_i\rho_i$, and sums are again over all components except the kinetic part of the scalar field.

The $\alpha_i(\phi)$ are the couplings of the dilaton with each component $i$, so they characterize the effect of the various components of the universe in the dynamics of the field. One may generically expect that the dilaton has different couplings to different components \cite{DPV2}. Experimental constraints impose a tiny coupling to baryonic matter, as we will discuss presently. In these models, this small coupling could naturally emerge due to a Damour-Polyakov type screening of the dilaton \cite{Polyakov}.

The relevant parameter here is the coupling of the dilaton field to hadronic matter. As discussed in \cite{Polyakov}, to a good approximation this is given by the logarithmic derivative of the QCD scale, since hadron masses are proportional to it (modulo small corrections). Assuming that all gauge fields couple, near the string cutoff, to the same $B_F(\phi)$, and in accordance with Eq. (\ref{eq:coupfunc}) which yields
\begin{equation}\label{bfhere}
B_F^{-1}(\phi)\propto (1-b_Fe^{-c\phi})\,,
\end{equation}
we can write
\begin{equation}\label{alphadefn}
\alpha_{had}(\phi)\sim 40 \frac{\partial\ln B_F^{-1}(\phi)}{\partial\phi}\,,
\end{equation}
(where the numerical coefficient is further described in \cite{DPV2}) and we finally obtain
\begin{equation}\label{alphahad}
\alpha_{had}(\phi)\sim 40\, b_F c\,e^{-c\phi}\,.
\end{equation}
Note that $c$ and $b_F$ are constant free parameters: the former one is expected to be of order unity and the latter one much smaller. Moreover, if we set $c=1$ (which we will do henceforth) we can also eliminate $b_F$ by writing 
\begin{equation}\label{alphahadrel}
\frac{\alpha_{had}(\phi)}{\alpha_{had,0}}=e^{-(\phi-\phi_0)}\,,
\end{equation}
(where $\phi_0$ is the value of the field today) and simultaneously writing the field equation in terms of $(\phi-\phi_0)$.

There are two local constraints. Firstly the Eddington parameter $\gamma$, which quantifies the amount of deflection of light by a gravitational source, has the value
\begin{equation}\label{eddingt}
\gamma-1=-2\alpha_{had,0}^2\,,
\end{equation}
and is constrained by the Cassini bound, $\gamma-1=(2.1\pm2.3)\times10^{-5}$ \cite{Cassini}.
Secondly the dimensionless E\"{o}tv\"{o}s parameter, quantifying violations to the Weak Equivalence Principle, has the value
\begin{equation}\label{eotvos}
\eta_{AB}\sim5.2\times10^{-5}\alpha_{had,0}^2\,,
\end{equation}
and recent torsion balance tests lead to $\eta_{AB}=(-0.7\pm1.3)\times10^{-13}$ \cite{Torsion}, while from lunar laser ranging one finds $\eta_{AB}=(-0.8\pm1.2)\times10^{-13}$ \cite{Lunar}. From these we conservatively obtain the bound
\begin{equation}\label{boundhad}
|\alpha_{had,0}|\le 10^{-4}\,.
\end{equation}
Using Eq. (\ref{alphahad}), and still assuming that $c\sim1$, this yields a bound on the product of $b_F$ and (the exponent of) $\phi_0$, namely $\phi_0\ge\ln{(|b_F|/2\times10^{-6})}$. Nevertheless, this is not explicitly needed: the evolution of the system will be determined by $\alpha_{had}$ rather than by $b_F$ or $\phi_0$.

These constraints do not apply to the dark sector (\textit{i.e.} dark matter and/or dark energy) whose couplings may be stronger. There are two possible scenarios to consider. A first possibility is that the dark sector couplings (which we will denote $\alpha_m$ and $\alpha_v$ for the dark matter and dark energy respectively) are also much smaller than unity, that is $\alpha_m,\alpha_v\ll1$. In this case the small field velocity leads to violations of the Equivalence Principle and variations of the fine-structure constant that are quite small. Indeed, for this case to be observationally realistic the fractions of the critical density of the universe in the kinetic and potential parts of the scalar field must be
\begin{equation}\label{lambdadil}
\Omega_k=\frac{1}{3}{\phi'}^2\ll1\,,\qquad \Omega_v\sim0.7;
\end{equation}
note that if one assumes a flat universe, then $\Omega_m+\Omega_k+\Omega_v=1$ (do not confuse the index $k$, which refers to the kinetic part of the scalar field, with the curvature term in standard cosmology, which we are setting to zero throughout). A more interesting possibility is that the dark couplings ($\alpha_m$ and/or $\alpha_v$) are of order unity. If so, violations of the Equivalence Principle and variations of the fine-structure constant are typically larger. In this case $\Omega_k$ may be more significant, and $\Omega_v$ should be correspondingly smaller \cite{quint}. Nevertheless the dark matter coupling is also constrained: during matter-domination the equation of state has the form
\begin{equation}
w_m(\phi)=\frac{1}{3} {\phi'}^2 \sim\frac{1}{3}\alpha_m^2\,.
\end{equation}

The present value of the field derivative is also constrained if one assumes a spatially flat universe; in that case the deceleration parameter
\begin{equation}\label{deccel}
q=-\frac{a{\ddot a}}{{\dot a}^2}=-1-\frac{\dot H}{H^2}\,
\end{equation}
can be written as
\begin{equation}\label{deccelhere}
{\phi_0'}^2=(1+q_0)-\frac{3}{2}\Omega_{m0} \,
\end{equation}
and using a reasonable upper limit for the deceleration parameter \cite{Neben} and a lower limit for the matter density (say, from the Planck mission \cite{XVI}) we obtain
\begin{equation}\label{phibound}
|\phi_0'|\le 0.3  \,,
\end{equation}
almost three times tighter than the one available at the time of \cite{DPV2}. Thus in this scenario both the hadronic coupling and the field speed today are constrained.

Moreover, we can use the field equation, Eq. (\ref{eq:field}), to set a consistency condition for $\phi_0'$. For this we only need to assume that the field is moving slowly today (a good approximation given the bounds on its speed) and therefore the $\phi''$ term should be subdominant in comparison with the other two. Then we easily obtain
\begin{equation}\label{phicons}
\phi_0'=\, - \, \frac{\alpha_{had}\Omega_b+\alpha_m\Omega_c+4\alpha_v\Omega_v}{\Omega_b+\Omega_c+2\Omega_v} \,,
\end{equation}
with all quantities being evaluated at redshift $z=0$. To avoid confusion we have denoted baryonic and cold dark matter by $\Omega_b$ and $\Omega_c$ respectively; naturally $\Omega_m=\Omega_b+\Omega_c$. We choose the cosmological parameters in agreement with recent Planck data \cite{XVI}, specifically setting the current fractions of baryons, dark matter and dark energy to be respectively $\Omega_{b}\sim0.04$, $\Omega_{c}\sim0.27$ and $\Omega_{\phi}=\Omega_{k}+\Omega_{v}\sim0.69$. Noting that $|\alpha_{had,0}|\le 10^{-4}$, that $|\phi_0'|\le 0.3$ and that $\Omega_k={\phi_0'}^2/3$ is necessarily small, we can consider three particular cases of this relation
\begin{itemize}
\item The {\bf dark coupling} case, where $\alpha_m=\alpha_v$ (and both are assumed to be constant), leads to
\begin{equation}\label{alphadark}
|\alpha_v|\lesssim 0.3 \frac{\Omega_m+2\Omega_v}{\Omega_c+4\Omega_v}\sim0.17 \,;
\end{equation}
\item The {\bf matter coupling} case, where $\alpha_m=\alpha_{had}$ (and both are field-dependent, as in Eq. \ref{alphahadrel}), leads to
\begin{equation}\label{alphamat}
|\alpha_v|\lesssim 0.3 \frac{\Omega_m+2\Omega_v}{4\Omega_v}\sim0.18 \,;
\end{equation}
\item The {\bf field coupling} case, where $\alpha_m=-\phi'$, leads to
\begin{equation}\label{alphafield}
|\alpha_v|\lesssim 0.3 \frac{\Omega_b+2\Omega_v}{4\Omega_v}\sim0.15 \,.
\end{equation}
\end{itemize}
Note that in all cases $\alpha_v$ is a constant (field-independent) parameter. Naturally these are back-of-the-envelope constraints that need to be improved by a more robust analysis, but they are enough to show that order unity couplings $\alpha_v$ will be strongly constrained. An additional constraint will come from atomic clock measurements, as we will now discuss.

\begin{figure}
\includegraphics[width=2.6in,keepaspectratio]{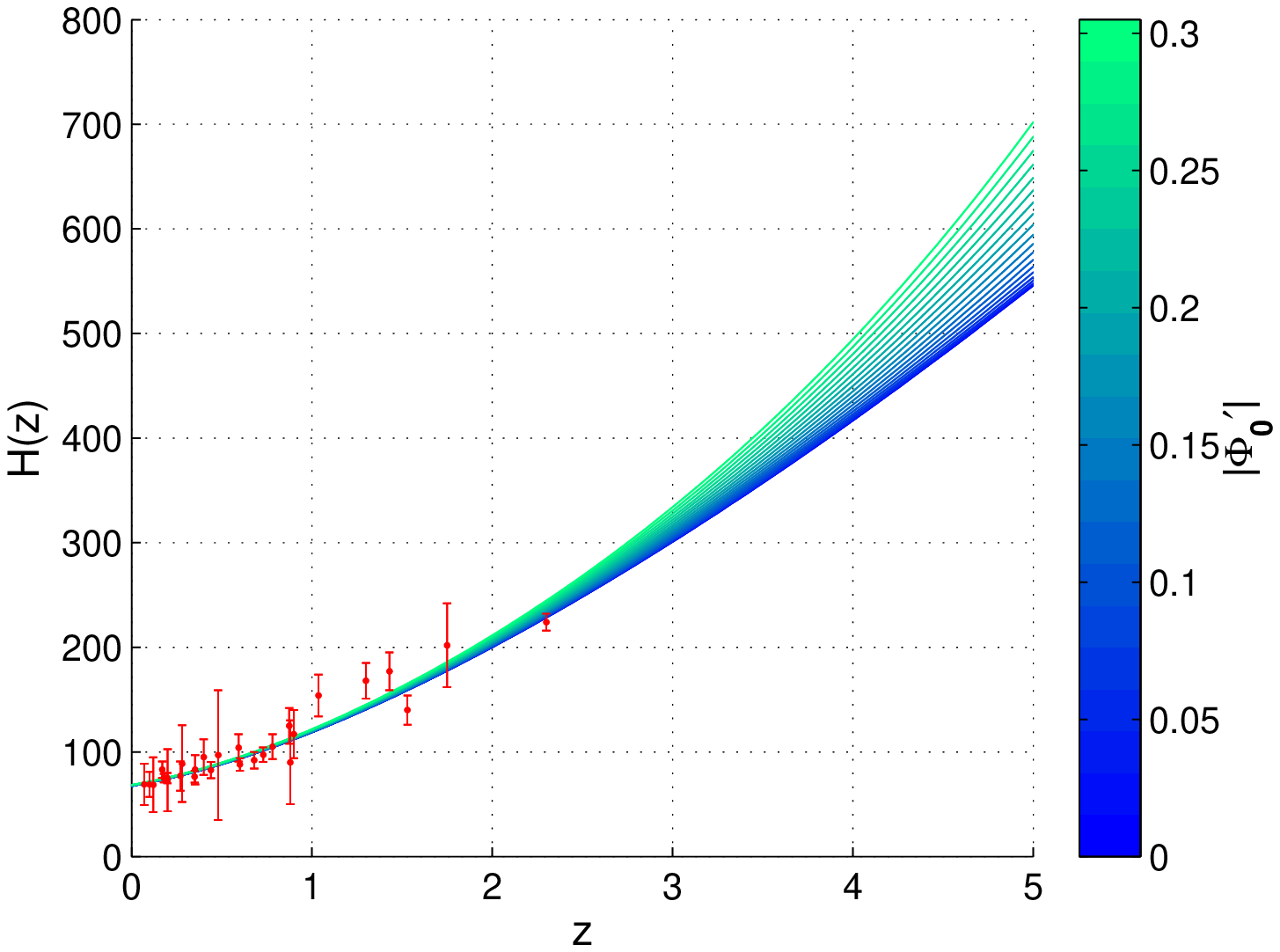}
\includegraphics[width=2.6in,keepaspectratio]{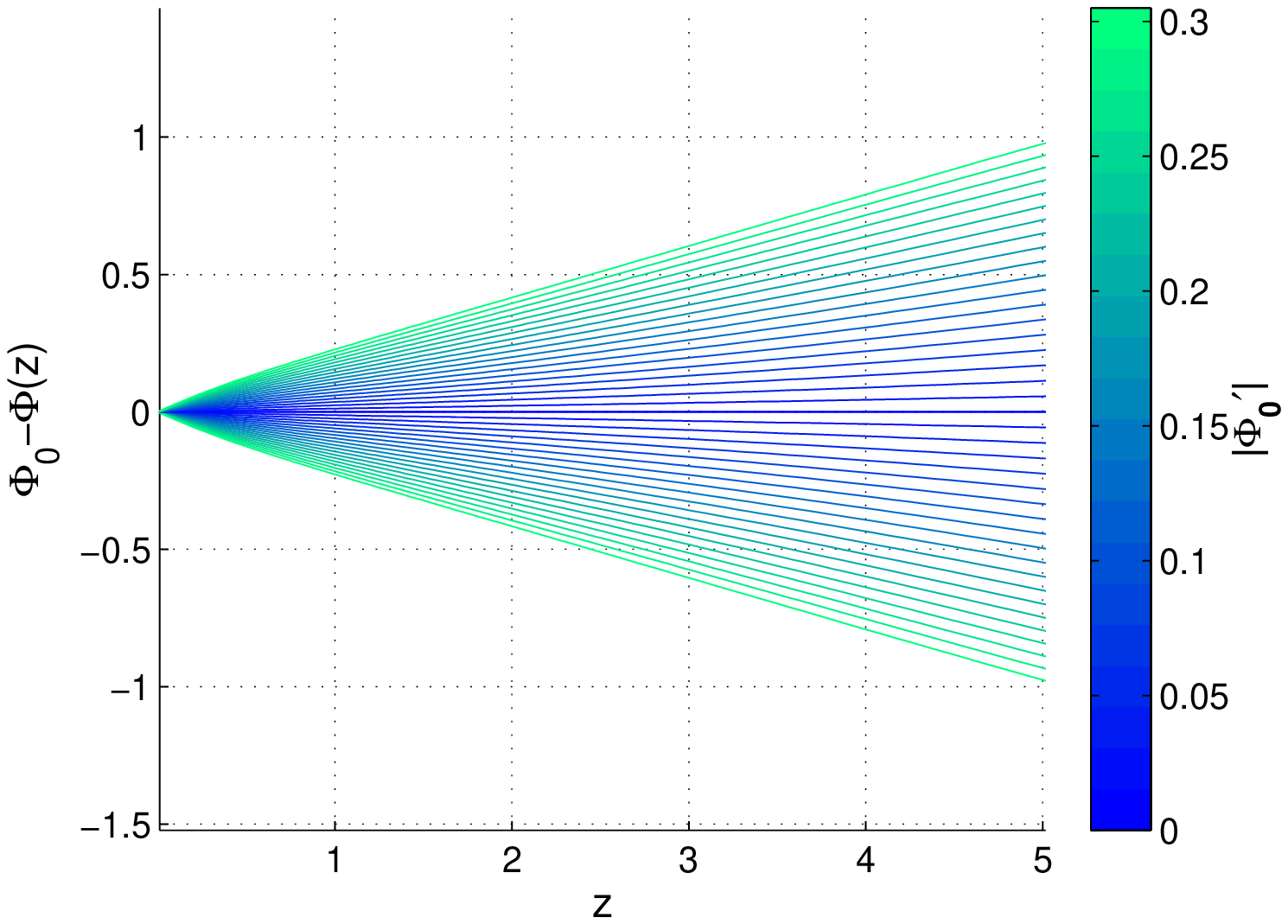}
\caption{Evolution of $H(z)$ in runaway dilaton models (left), compared to the measurements in \cite{Farooq}, for $|\alpha_{had,0}|=10^{-4}$ and $\phi_0'$ spanning the observationally allowed range. The right-side panel depicts the evolution of the field for the same parameter choices.}
\label{fig1}
\end{figure}

\section{Varying fine-structure constant}

Given some recent evidence, from archival Keck and VLT data, of space-time variations of the fine-structure constant $\alpha$ \cite{Webb}, it's interesting to study its behavior in this class of models. Consistently with our previous assumption that all gauge fields couple to the same $B_F$, here $\alpha$ will be proportional to $B_F^{-1}(\phi)$, as given by Eq. (\ref{bfhere}). Note that this will also imply that $\alpha$ will be related to the hadronic coupling, as further discussed below.

The original work of Damour \textit{et al.} \cite{DPV2} shows (under the same assumptions as we are using here) that the evolution of $\alpha$ is given by
\begin{equation}\label{alphazero}
\frac{1}{H}\frac{\dot\alpha}{\alpha}=\frac{b_Fce^{-c\phi}}{1-b_Fce^{-c\phi}}\,{\phi'}\sim b_Fce^{-c\phi}{\phi'}\sim\frac{\alpha_{had}}{40}{\phi'}\,.
\end{equation}
In particular this equation applies at the present day (describing the current running of $\alpha$) and this variation is constrained by the Rosenband bound \cite{Rosenband}
\begin{equation}
\left(\frac{1}{\alpha}\frac{{\rm d}\alpha}{{\rm d}t}\right)_0=(-1.6\pm2.3)\times 10^{-17}{\rm yr}^{-1}\,;
\end{equation}
assuming the Planck value for the Hubble constant $H_0=(67.4\pm1.4)\, {\rm km/s/Mpc}$, we find
\begin{equation}\label{jointphibound}
|\alpha_{had,0}{\phi'}_0| \sim |b_Fce^{-c\phi_0}{\phi'}_0| \le 3\times 10^{-5}\,.
\end{equation}
Thus atomic clock experiments constrain the product of the hadronic coupling and the field speed today. It is interesting to note that this constraint---which stems from microphysics---is comparable to the one obtained by multiplying the individual constraints on each of them, which are given respectively by Eq. \ref{boundhad} and Eq. \ref{phibound} and come from macrophysics (Solar System or torsion balance tests, plus a cosmology bound).

\begin{figure}
\includegraphics[width=2.6in,keepaspectratio]{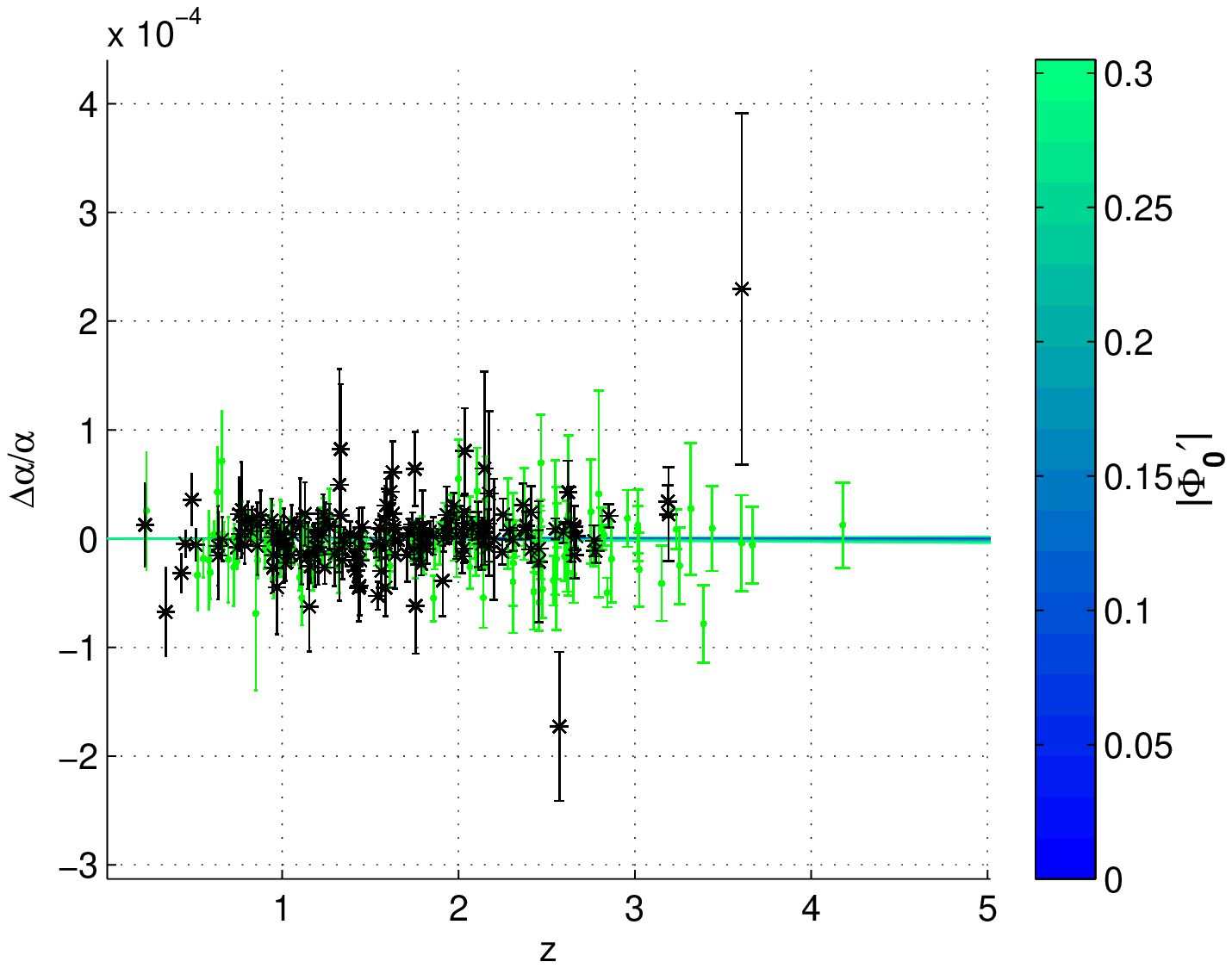}
\includegraphics[width=2.6in,keepaspectratio]{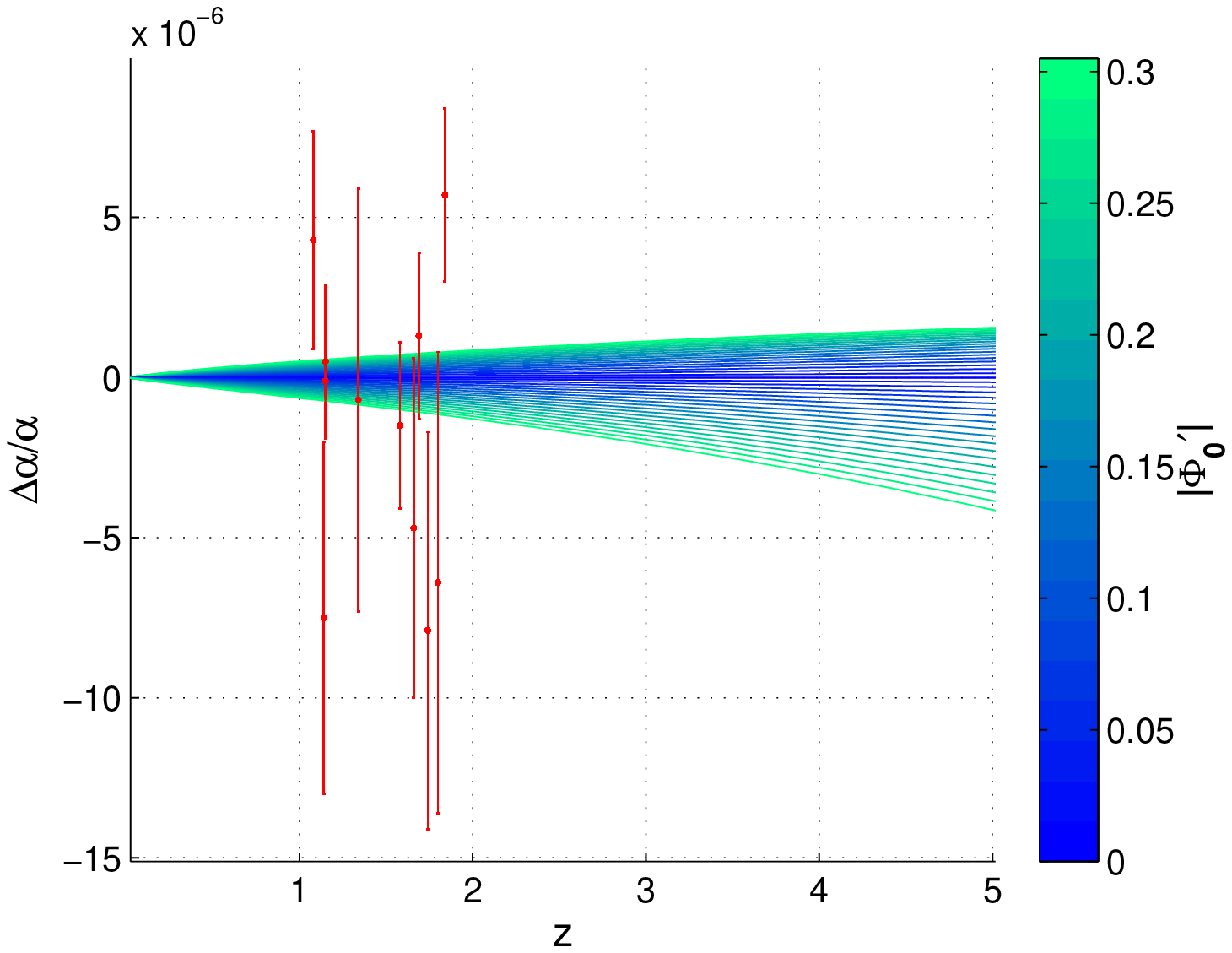}
\caption{Evolution of $\alpha$, plotted with the same conventions as in Fig. \protect\ref{fig1}. The data of \protect\cite{Webb} is plotted in the left panel (VLT data as black asterisks, Keck data as green circles) while the data of Table \protect\ref{tablealpha} is shown as red circles in the right panel. One-sigma uncertainties are shown in all cases.}
\label{fig2}
\end{figure}

In \cite{DPV2} the authors obtain approximate solutions for the evolution of $\alpha$ by assuming that $\phi'=const.$ in both the matter and the dark energy eras (naturally the two constants are different). However, by integrating Eq. (\ref{alphazero}) or by directly using the relation between $\alpha$ and $B_F(\phi)$ we can express the redshift dependence of $\alpha$ in the general form
\begin{equation}\label{evolfull0}
\frac{\Delta\alpha}{\alpha}(z)\equiv\frac{\alpha(z)-\alpha_0}{\alpha_0}=B_F^{-1}(\phi(z))-1 =b_F\left(e^{-\phi_0}-e^{-\phi(z)}\right)\,,
\end{equation}
where for simplicity we have again set $c\sim1$. This can also be recast in the more suggestive form
\begin{equation}\label{evolfull}
\frac{\Delta\alpha}{\alpha}(z)=\frac{1}{40}\alpha_{had,0}\left[1-e^{-(\phi(z)-\phi_0)}\right]\,.
\end{equation}
Thus the behaviour of $\Delta\alpha/\alpha$ close to the present day depends both on $\alpha_{had,0}$ (which provides an overall normalization) and on the speed of the field, ${\phi_0'}$, which can also be related to the values of the couplings as in Eq. (\ref{phicons}).

In our analysis we will use both the data of Webb {\it et al.} \cite{Webb} (which is a large dataset of archival data measurements) and the smaller and more recent dataset of dedicated measurements listed in Table \ref{tablealpha}. The latter include the recent first result of the UVES Large Program for Testing Fundamental Physics \cite{LP1,LP3}, which is expected to be the one with a better control of possible systematics. The source of the data in this Table is also further discussed in \cite{Ferreira}. We emphasize that all the data we use comes from high-resolution spectroscopy comparisons of optical/UV fine-structure atomic doublets, which are only sensitive to the value of $\alpha$---and not, say, to the values of particle masses (ratios of which can be probed by other means) \cite{Kozlov}.

\begin{table}
\begin{center}
\begin{tabular}{|c|c|c|c|c|}
\hline
 Object & z & ${ \Delta\alpha}/{\alpha}$ (ppm) & Spectrograph & Ref. \\ 
\hline\hline
3 sources & 1.08 & $4.3\pm3.4$ & HIRES & \protect\cite{Songaila} \\
\hline
HS1549$+$1919 & 1.14 & $-7.5\pm5.5$ & UVES/HIRES/HDS & \protect\cite{LP3} \\
\hline
HE0515$-$4414 & 1.15 & $-0.1\pm1.8$ & UVES & \protect\cite{alphaMolaro} \\
\hline
HE0515$-$4414 & 1.15 & $0.5\pm2.4$ & HARPS/UVES & \protect\cite{alphaChand} \\
\hline
HS1549$+$1919 & 1.34 & $-0.7\pm6.6$ & UVES/HIRES/HDS & \protect\cite{LP3} \\
\hline
HE0001$-$2340 & 1.58 & $-1.5\pm2.6$ &  UVES & \protect\cite{alphaAgafonova}\\
\hline
HE1104$-$1805A & 1.66 & $-4.7\pm5.3$ & HIRES & \protect\cite{Songaila} \\
\hline
HE2217$-$2818 & 1.69 & $1.3\pm2.6$ &  UVES & \protect\cite{LP1}\\
\hline
HS1946$+$7658 & 1.74 & $-7.9\pm6.2$ & HIRES & \protect\cite{Songaila} \\
\hline
HS1549$+$1919 & 1.80 & $-6.4\pm7.2$ & UVES/HIRES/HDS & \protect\cite{LP3} \\
\hline
Q1101$-$264 & 1.84 & $5.7\pm2.7$ &  UVES & \protect\cite{alphaMolaro}\\
\hline
\end{tabular}
\caption{\label{tablealpha}Recent dedicated measurements of $\alpha$. Listed are, respectively, the object along each line of sight, the redshift of the measurement, the measurement itself (in parts per million), the spectrograph, and the original reference. The recent UVES Large Program measurements are \cite{LP1,LP3}. The first measurement is the weighted average from 8 absorbers in the redshift range $0.73<z<1.53$ along the lines of sight of HE1104-1805A, HS1700+6416 and HS1946+7658, reported in \cite{Songaila} without the values for individual systems. The UVES, HARPS, HIRES and HDS spectrographs are respectively in the VLT, ESO 3.6m, Keck and Subaru telescopes.}
\end{center}
\end{table}

Note that since in the current work we will be interested in the evolution of $\alpha$ at relatively low redshifts, one could think of linearizing the field evolution
\begin{equation}\label{evollinear}
\phi\sim\phi_0+{\phi_0'}\ln{a}\,,
\end{equation}
in which case Eq. (\ref{evolfull}) takes the simpler form
\begin{equation}\label{evolslow}
\frac{\Delta\alpha}{\alpha}(z)\approx\, -\frac{1}{40}\alpha_{had,0} {\phi_0'}\ln{(1+z)}\,;
\end{equation}
this is indeed what is obtained with the simplifying assumptions of \cite{DPV1,DPV2}. Nevertheless, as shown in the second panel of Fig. \ref{fig1}, $\phi-\phi_0$ can still be of order unity by redshift $z=5$ for values of the coupling that saturate the current bounds, and therefore in what follows the evolution of $\alpha$ will be calculated using the full equations.

\section{Current constraints}

By numerically solving the previously discussed Friedmann and scalar field equations we can study the cosmological dynamics of this model. We will start by assuming that the value of $\alpha_{had,0}$ is the maximal one allowed by Eq. (\ref{alphahad})---we will relax this assumption later on. We allow $\phi_0'$ to vary in the whole range allowed by Eq. (\ref{phibound}), and we further assume the dark coupling case, where $\alpha_m=\alpha_v$; it then follows from in Eq. (\ref{phicons}) that $\phi_0'\approx -1.79\alpha_v$.

\begin{figure}
\includegraphics[width=2.6in,keepaspectratio]{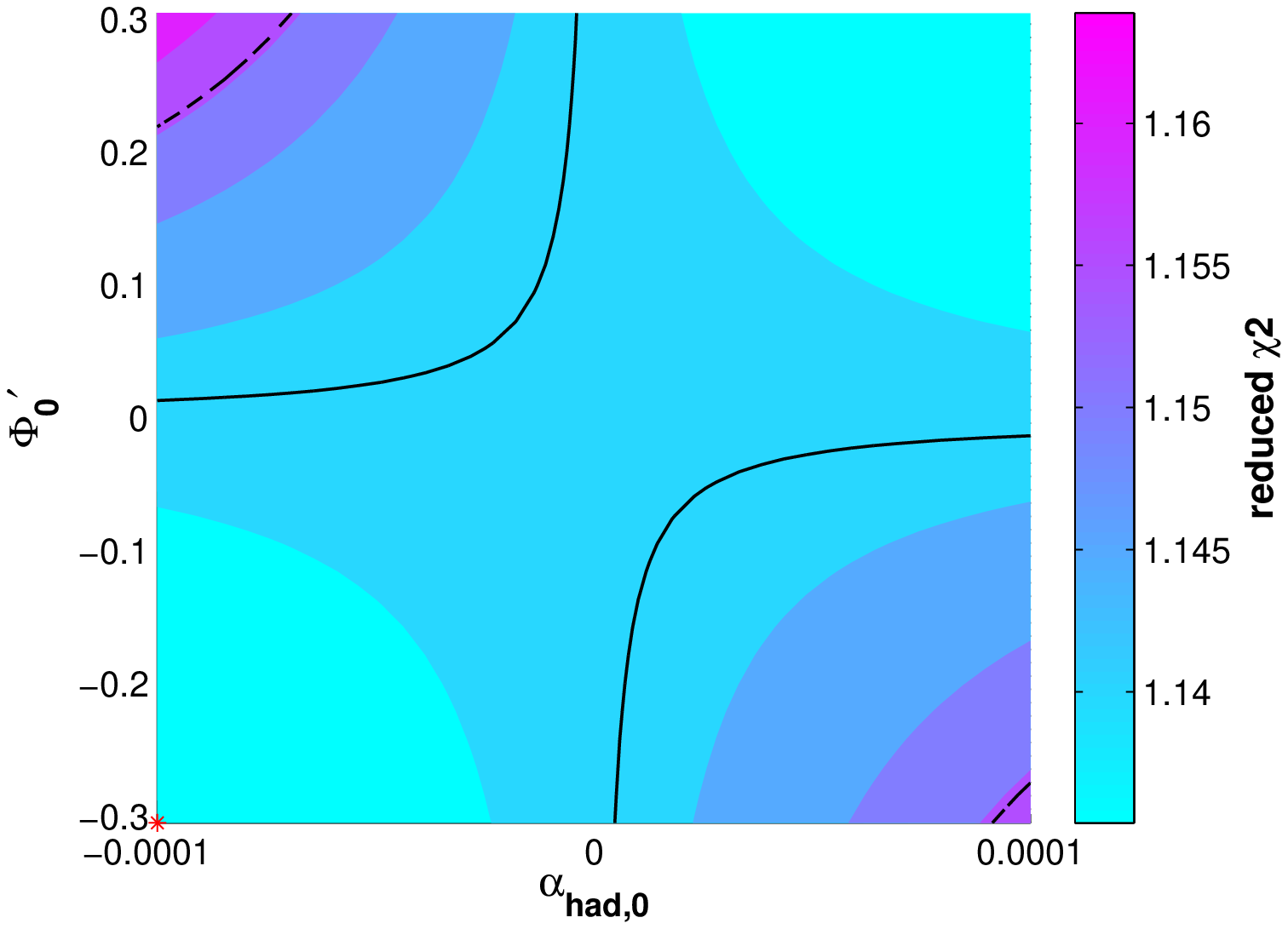}
\includegraphics[width=2.6in,keepaspectratio]{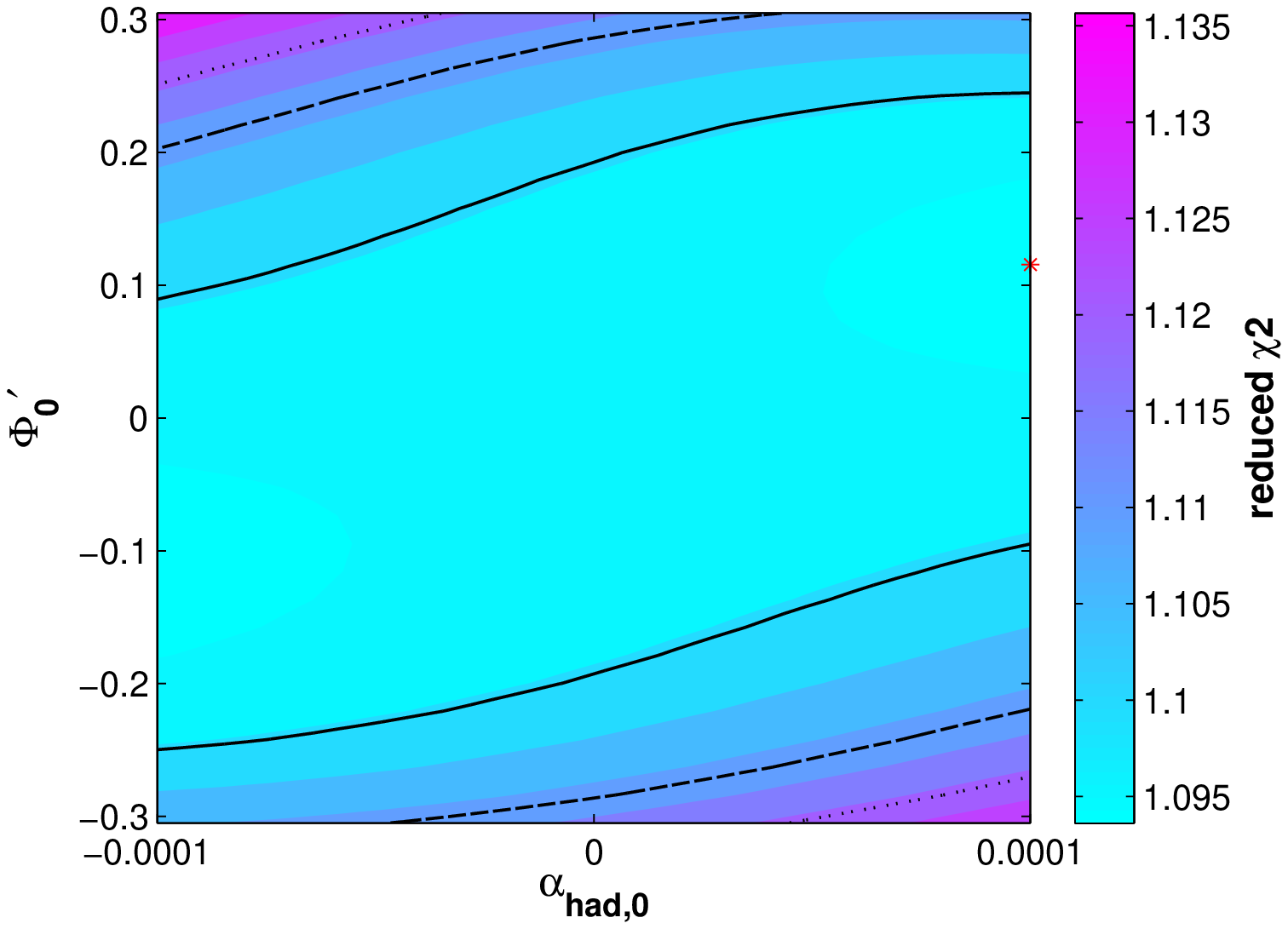}
\caption{Constraints in the $\alpha_{had,0}$--$\phi_0'$ space, with other parameters as described in the text; in the left panel only the $\alpha$ measurements were used, while the right one the $H(z)$ measurements were also included. The colormap shows the reduced chi-square, while the black solid, dashed and dotted lines identify the one, two and three-sigma confidence regions in this parameter space.}
\label{fig3}
\end{figure}

We choose the same cosmological parameters as previously discussed. Note that in this model the dark energy equation of state is
\begin{equation}\label{darkenergyeos}
1+w_0=\frac{2\Omega_{k}}{\Omega_{k}+\Omega_{v}}=\frac{2}{3}\frac{{\phi_0'}^2}{\Omega_{k}+\Omega_{v}}\,,
\end{equation}
and the range of allowed values for ${\phi_0'}$ (specifically, $|\phi_0'|\le 0.3$) leads to $-1\le w_0<-0.91$, which is perfectly compatible with current observational bounds \cite{XVI}. We then numerically integrate the dynamical equations of this model backwards in time. The evolution of the Hubble parameter for this set of models is plotted in Fig. \ref{fig1}, and compared to the available measurements, as compiled in \cite{Farooq}. As expected the sign of the coupling $\alpha_{had,0}$ has a negligible effect on $H(z)$ (since the coupling itself is very small), while that of the field speed is more noticeable.

We then calculate the evolution of $\alpha$ in these models; this is shown in Fig. \ref{fig2}, again for the maximally allowed $|\alpha_{had,0}|=10^{-4}$. With these parameter choices the typical variations are at the parts per million level, comparable to the sensitivity of the current measurements \cite{LP1,LP3}. The value of $\alpha$ also depends on the present speed of the field (and not only on its absolute value), which can be understood from Eq. \ref{evolfull}.

As a second step in our analysis, we now relax the assumption of $\alpha_{had,0}$ fixed to its maximum allowed value and let it vary freely. We use the available data to constrain it, together with the field speed. The results of this analysis are shown in Fig. \ref{fig3}. Using all available $\alpha$ data (both that of \cite{Webb} and the dedicated measurements of Table \ref{tablealpha}) one finds no significant evidence for a non-zero coupling $\alpha_{had,0}$. While the weighted mean of the data in Table \ref{tablealpha} is consistent with no variations, that of \cite{Webb} is slightly negative; this explains why in the first panel of Fig. \ref{fig3} there is a slight preference for similar signs for the field speed and the coupling (however, this is not statistically significant). We thus see that with the $\alpha$ data alone the constraints are not that much stronger than we already discussed above. The addition of Hubble parameter measurements does constrain the current speed of the field to be small, and the combination of the two datasets yields the constraints in the second panel of Fig. \ref{fig3}. In both cases the model is compatible with the current data.

We caution the reader that this analysis assumed fixed values of the cosmological parameters $\Omega_{b}$, $\Omega_{c}$ and $\Omega_{\phi}$, but we expect the results not to change significantly had we allowed them to vary and marginalized over them. Perhaps more relevant are our `maximal' assumptions for the dark sector couplings, which can be justified in the context of a preliminary assessment of the feasibility of the model. Thus our present results suggest that a more thorough exploration of this parameter space is justified, but we leave it for a more detailed follow-up publication.

\begin{figure}
\includegraphics[width=3in,keepaspectratio]{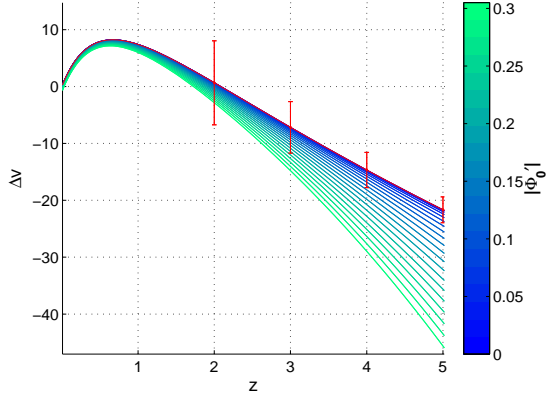}
\caption{Redshift drift signal for allowed runaway dilaton models (plotted with the same conventions as in Fig. \protect\ref{fig1}) compared to the standard $\Lambda$CDM model (red curve, shown with errorbars expected for E-ELT measurements).}
\label{fig4}
\end{figure}

\section{Outlook}

While current astrophysical and laboratory constraints on $\alpha$ provide (together with Equivalence Principle tests) interesting constraints on string theory inspiredscenarios, prospects for further improvements are excellent in the context of European Extremely Large Telescope (E-ELT): this will not only enable much more sensitive measurements of the fine-structure constant but it will also open a new and complementary observational window into these models.

Redshifts of cosmologically distant objects drift slowly with time \cite{Sandage}. This provides a direct measurement of the Universe's expansion history, with the advantage of being a non-geometric, completely model-independent test, uniquely probing the global dynamics of the metric \cite{Liske}. Rather than mapping our past light-cone, it directly measures evolution by comparing past light-cones at different times.  While plans are being developed to carry out these measurements at low redshift (with the SKA \cite{SKA} and intensity mapping experiments \cite{Yu}), the E-ELT offers the unique advantage of probing deep in the matter era and thus a much larger redshift lever arm. The precision needed for these measurements, a few cm/s, will be reached with the E-ELT's high-resolution ultra-stable spectrograph currently dubbed ELT-HIRES. A Phase A study \cite{Liske} led to the following estimate for the spectroscopic velocity precision
\begin{equation}\label{eq:hiresprec}
\sigma_v=1.35\left(\frac{S/N}{2370}\right)^{-1}\left(\frac{N_{QSO}}{30}\right)^{-1/2}\left(\frac{1+z_{QSO}}{5}\right)^{-1.7}\,;
\end{equation}
this depends on the signal-to-noise of the spectra, as well as on the number and the redshift of the quasar absorption systems used. The signal for a given model can be derived from the definition of redshift and expressed in a model independent way in terms of the spectroscopic velocity (which is the actual observable) as
\begin{equation}\label{eq:slsignal}
\frac{\Delta v}{c}=\frac{\Delta t}{(1+z)} \left[H_0(1+z)-H(z)\right]\,,
\end{equation}
where $\Delta t$ is the timespan of the measurements.

The drift signal for our range of models is plotted in Fig. \ref{fig4} and compared to $\Lambda$CDM, for $\Delta t=30$ years. The error bars depict the expected accuracy of ELT- HIRES, assuming 40 sources with $S/N=2000$. As with several other alternatives to $\Lambda$CDM studied in the literature \cite{Quercellini}, it is clear that the drift signal in runaway dilaton models can differ significantly from that of $\Lambda$CDM, and ELT-HIRES will thus be able to distinguish the two paradigms and set tighter constraints both on $\alpha_{had,0}$ and on the dark sector couplings.

In conclusion, the runaway dilaton scenario is compatible with current data. It (and many other models) will be subject to much more stringent tests as the next generation of high-resolution, ultra-stable spectrographs becomes available. A roadmap for these tests is further discussed in \cite{grg}. Meanwhile, the E\"{o}tv\"{o}s parameter sensitivity is also expected to improve to $2\times10^{-15}$ with a hypothetical STE-QUEST \cite{stequest} and to $10^{-18}$ with STEP \cite{step}, and these will provide complementary constraints. Thus quantitative astrophysical tests of string-inspired scenarios will soon become possible.

\section*{Acknowledgements}
This work was done in the context of project PTDC/FIS/111725/2009 (FCT, Portugal). CJM is also supported by an FCT Research Professorship, contract reference IF/00064/2012, funded by FCT/MCTES (Portugal) and POPH/FSE. MM acknowledges the DFG TransRegio TRR33 grant on The Dark Universe. EC acknowledges funding from ERC grant 259505. The Dark Cosmology Centre is funded by the Danish National Research Foundation.

\bibliographystyle{model1-num-names}
\bibliography{dilaton}

\end{document}